# Spectrally resolved single-photon imaging with hybrid superconducting - nanophotonic circuits

O. Kahl[1,2], S. Ferrari[1,2], V. Kovalyuk[2,3], A. Vetter[2], G. Lewes-Malandrakis[4], C. Nebel[4], A. Korneev[3,5], G. Goltsman[3,6], W. Pernice[1]*

[1]University of Münster, Institute of Physics, Heisenbergstr. 11, 48149 Münster, Germany.

[2]Karlsruhe Institute of Technology (KIT), Institute of Nanotechnology, Hermann-von-Helmholtz-Platz 1, 76344 Eggenstein-Leopoldshafen, Germany.

[3]Department of Physics, Moscow State Pedagogical University, Moscow 119992, Russia.

[4]Fraunhofer Institute for Applied Solid State Physics, Tullastr. 72, 79108 Freiburg, Germany.

[5]Moscow Institute of Physics and Technology (State University), Moscow 141700, Russia.

[6] National Research University Higher School of Economics, 20 Myasnitskaya Ulitsa, 101000, Russia

*Correspondence to:  wolfram.pernice@uni-muenster.de

The detection of individual photons is an inherently binary mechanism, revealing either their absence or presence while concealing their spectral information. For multi-color imaging techniques, such as single photon spectroscopy, fluorescence resonance energy transfer microscopy and fluorescence correlation spectroscopy, wavelength discrimination is essential and mandates spectral separation prior to detection. Here, we adopt an approach borrowed from quantum photonic integration to realize a compact and scalable waveguide-integrated single-photon spectrometer capable of parallel detection on multiple wavelength channels, with temporal resolution below 50 ps and dark count rates below 10 Hz. We demonstrate multi-detector devices for telecommunication and visible wavelengths and showcase their performance by imaging silicon vacancy color centers in diamond nanoclusters. The fully integrated hybrid superconducting-nanophotonic circuits enable simultaneous spectroscopy and lifetime mapping for correlative imaging and provide the ingredients for quantum wavelength division multiplexing on a chip.

# Introduction

Photonic quantum technologies provide next-generation tools for the implementation of quantum information processing schemes using classical nanophotonic circuitry (*1–8*). Operating waveguide based systems with single-photon input grants access to the established fabrication and design techniques used for integrated optics and enables circuit complexity which is out of reach for table-top, free-space implementations. Current research efforts in optical quantum information processing pursue the full integration of active quantum optical components – single-photon sources (*9–12*) and single-photon detectors (*7, 8, 13–16*) – with reconfigurable photonic circuitry in a single chip. This development holds potential to resolve current limitations in terms of stability, robustness and component scalability. While the joint integration of all relevant components on a chip is yet to be achieved, the emergence of highly performant waveguide-integrated single-photon detectors (*13, 17–20*) opens up additional avenues beyond applications in quantum optics. Here, we demonstrate how the thriving fields of optical sensing and imaging benefit from nanophotonic integration through the use of such detectors.

State-of-the-art microscopy techniques such as fluorescence resonance energy transfer (FRET) or fluorescence lifetime imaging (FLIM) are indispensable tools in the modern life sciences, allowing molecular networks and intracellular activities to be monitored with great spatial and temporal precision (*21–27*). Data acquisition often requires the detection of weak optical signals down to the single-photon level and necessitates highly sensitive detectors with broad spectral range (*28, 29*). Standard experimental and commercial implementations comprise single-photon detectors, wavelength-selective optics for spectroscopic analysis and fast time-correlated single-photon counting (TCSPC) electronics for correlative investigation. For unperturbed, accurate measurements over extended time periods, these applications rely on the single-photon detectors' high efficiency, low noise level, and precise timing resolution, as well as the robust, immovable installation of all optical elements. Current state-of-the-art single-photon detectors suffer from performance-degrading effects such as after-pulsing or low efficiency and large timing jitter, particularly in the near-infrared spectral range (*30*, *31*). In addition, the bulk optical components typically used in classical setups are prone to misalignment. Most experimental installations are therefore subject to stability constraints and offer only limited scalability. Photonic integration constitutes a promising resolution to these issues. By combining nanophotonic integrated circuits with superconducting single photon detectors we implement a scalable single photon spectrometer which provides single photon resolution on multiple wavelength channels and is inherently stable.

## Materials and Methods

### Integrated single-photon spectrometer concept and design

Our spectrometer concept is based on the use of high performance waveguide-integrated superconducting nanowire single-photon detectors (SNSPDs) in conjunction with wavelength-discriminating integrated photonic circuitry. SNSPDs provide significant advantages over current state-of-the-art single-photon detectors: they offer superb detection efficiency and impeccable timing characteristics over a wide spectral domain, encompassing visible and infrared wavelengths (*13*, *17*, *32*, *33*). In addition, their inherently integrated design enables the seamless conflation of multiple SNSPDs with advanced on-chip photonic circuitry while maintaining the detectors' efficiency and temporal precision. The co-integration of multiple SNSPDs with wavelength-separating photonic circuitry therefore offers a convenient approach to on-chip single-photon spectroscopy with high timing accuracy and fast data acquisition rates. Beside pure spectroscopic analysis at the single-photon level, the concept enables advanced techniques such as FLIM and FRET or its exotic siblings multi-color and N-way FRET (*34*, *35*) in a single device.

We realize fully integrated single-photon spectrometers (SPSs) by co-integrating eight SNSPDs with an arrayed waveguide grating (AWG) on a silicon-nitride ($Si_3N_4$) on insulator substrate (Fig. 1). The chosen platform enables broadband optical operation, filtering and multiplexing combined with highly efficient on-chip single-photon detection. The fabrication is based on a top-down approach which allows for the reliable and reproducible integration of all active and passive photonic components, as described in the Methods section and the supplementary material.

Light is coupled into the on-chip circuitry through a focusing grating coupler. The AWG spatially separates the broadband optical input signal and routes individual spectral components towards the SNSPD array for photon detection. Because the AWG distributes different wavelengths into different waveguides, the design is directly compatible with waveguide integrated SNSPDs. The count rates measured by the individual detectors allow for the reconstruction of the optical input spectrum. The utilization of an AWG leaves the temporal photon distribution unaffected which allows us to exploit the SNSPDs' high timing accuracy and thus expand the conventional spectroscopic concept by a temporal dimension.

**Device fabrication**

The two sets of hybrid superconducting-nanophotonic chips are fabricated using multi-step electron-beam (e-beam) lithography with subsequent dry etching and thin film deposition steps. Semiconductor multi-layer structures consisting of silicon-nitride ($Si_3N_4$) and silicon-dioxide ($SiO_2$) thin films on top of a silicon carrier wafer are used as starting materials. Different layer thickness configurations are used for the chips for 1550 nm and 738 nm optical wavelength, respectively: the 1550 nm design employs 450 nm $Si_3N_4$ and 2.6 µm thick $SiO_2$ layers, while the 738 nm design is based on 200 nm thick $Si_3N_4$ and 2.0 µm thick $SiO_2$ layers.

The template wafers are diced into 15x15mm$^2$ pieces for mounting into the e-beam exposure system (50kV JEOL 5500). In a first step, a 4 nm thin niobium-nitride (NbN) layer is deposited by magnetron-sputtering in an argon-nitrogen atmosphere (*40*). Subsequently, gold contact pads and alignment markers are realized by electron beam (e-beam) lithography using positive tone polymethyl methacrylate (PMMA) 8.0 e-beam resist. The resist is developed in a 1:3 methyl-isobutyl-ketone isopropanol (MIBK) solution and 5 nm chromium and 120 nm gold layers are vapor deposited by a custom-made e-beam physical vapor deposition (PVD) machine. The lift-off is performed in acetone with mild sonication. The NbN nanowire detector elements are structured in a second e-beam lithography step using hydrogen-silsesquioxane (HSQ) as negative tone resist. To facilitate HSQ adhesion a 5 nm thick SiO2 layer is vapor-deposited via PVD prior to resist application. Development of the exposed sample in a 6.25% tetramethylammonium hydroxide (TMAH) bath to produce a solid etch mask is followed by a carefully timed reactive ion etching (RIE) step in a tetrafluoromethane ($CF_4$) atmosphere to transfer the nanowire pattern into the NbN layer. Finally, the photonic circuit structures are fabricated using ma-N 2403 as negative tone resist in a final e-beam lithography step. The written sample is developed in a MF319 solution and the structures are etched using a trifluoromethane ($CHF_3$) chemistry. Residual resist after etching is removed in an N-methyl-2-poyrrolidone (NMP) bath.

On each fabricated sample several spectrometer devices including eight SNSPDs on the output waveguides as well as several single-detector reference devices are realized. The latter allow for additional detector characterization possibilities on the same chip and help confirm the data obtained at the spectrometer devices.

**Optical SPS characterization at room temperature**

The arrayed waveguide grating (AWG) spectrometers' performance is characterized by inserting and extracting light into the on-chip circuitry through focusing grating couplers and an array of fibers properly positioned above the couplers. A computer-controlled 4D piezo translation-rotation stage is used to facilitate alignment (see supplementary Fig. S1). The setup allowed us to investigate multiple photonic circuits on the same chip after nanofabrication. Light from a fiber-coupled laser (tunable NIR laser New Focus TSL-6600 for 1550nm) or a white light continuum source (Leukos SM-30-UV to cover the wavelength range around 738 nm) is routed toward the input coupler. A fiber switch (Dicon GP700) is used for output channel selection. A power meter for the near-IR wavelength range (HP 8163A) and a modular spectrometer for the visible wavelength range (Ocean Optics JAZ) measure the transmitted powers in the 1500 nm and 738 nm spectral regions, respectively.

**Cryogenic measurement setup**

A liquid helium flow cryostat is used to cool the photonic chip containing both the AWGs and the detectors down to a stable base temperature of 1.6 K and $10^{-5}$ mbar pressure inside the sample chamber. The photonic chip is mounted on a 4D rotation-translation stage (Attocube Systems) inside the cryostat's sample chamber underneath an array of optical fibers and a multi-RF contact probe (Cascade Microtech). Optical connection is established by positioning the photonic circuits' grating couplers directly below the fibers and the detectors are connected by bringing the on-chip contact pads in physical contact with the RF probe. The cryostat provides 8 RF output lines for measuring each of the 8 output channels of the AWG device. Each superconducting single photon detector (SNSPD) is connected to a low-noise current source and read-out circuitry. The on-chip detector is current-biased by applying a constant voltage (Keithley 2400) over a 1 M$\Omega$ resistor via a high-frequency bias-tee (Mini-Circuits ZFBT-6GW+). High frequency components of the bias current are filtered using a low-pass filter (BLP-5, DC 5MHz). The voltage pulse generated upon detection of a photon is amplified by low-noise RF amplifiers (ZFL-1000LN+, +50 dB gain), connected to a SSPD over the RF channel of a bias-tee. Arriving voltage pulses are registered either with a time-correlated single photon counter (TCSPC, Picoharp 300, Picoquant), a fast sampling oscilloscope (Infinium 6 GHz, Agilent), or a pulse counter (Agilent 53132A).

## Results and discussion

### Spectrometer characterization

Two circuit variants are fabricated: one implementation is designed for operation in the near-infrared (NIR) at telecom wavelengths around 1550 nm, the other targets the red visible regime around 738 nm. These wavelength regions thus cover the telecommunication C-band used for optical fiber communication and quantum communication, and tangent the visible wavelength region where imaging applications in the life sciences are carried out. Both hybrid nanophotonic chips are realized through a multi-step electron-beam lithography process with subsequent dry etching and thin film deposition steps, as described in detail in the Methods section and the supplementary material. We characterize the integrated SPS in the telecommunications range and demonstrate the device's imaging and sensing capabilities by spectrally and temporally analyzing the fluorescence light emitted from silicon vacancy (SiV) color centers at 738 nm in diamond. In combination with a custom confocal scan head (Fig. 1a) we determine spatially resolved lifetime maps of the diamond nanocrystal sample.

First, we assess the optical transmission through the devices at room temperature using a custom multi-port measurement system (see Methods section and the supplementary material). For this purpose, each output channel of the AWG is furnished with an additional grating coupler for light extraction (see Fig. 1d). The transmission spectrum measured at the NIR device is depicted in Fig. 2a. It shows eight separate transmission channels with low cross-talk of -18dB. The wavelength channels are designed to fall within the telecommunication C-band and feature a channel spacing of 2.2 nm. The chip is then mounted inside a cryogenic measurement system to analyze the spectrometer performance at low temperature using the on-chip single photon detectors. The temperature is stabilized at 1.6K well below the transition temperature of the superconducting nanowires. The automated setup allows us to access and measure multiple spectrometer devices optically and electrically on multiple input channels. We define the NIR device's overall on-chip detection efficiency $\eta = 1/8 \sum_{i=1}^{8} \eta^{(i)}$ with contributions $\eta^{(i)}$ of the detectors $i \in \{1, ...,8\}$. Individual efficiencies $\eta^{(i)}$ are extracted by comparing a well-calibrated photon flux of $\Phi = 10^6 \, s^{-1}$ inside the input waveguide to the measured detector count rates $R_c^{(i)}$ after correcting for the detectors' dark count rates $R_{dc}^{(i)}$, i.e. $\eta^{(i)} = \left[R_c^{(i)} - R_{dc}^{(i)}\right]/\Phi$. The detectors' bias current $I_b$ is slowly raised from 50% to 90% of their respective critical currents $I_c$ and the input wavelength is swept across the device's free spectral range from 1532.5 nm to 1557.5 nm. Fig. 2b shows the resulting efficiency plot which clearly exhibits the eight transmission windows in which the individual detectors are able to detect single-photons. The efficiency data agrees well with the optical transmission data obtained at room temperature (Fig. 2a). With the bias current set to 90% of the critical current (Fig. 2c) an overall SPS device efficiency of $\eta = (18.93 \pm 6.21)\%$ is obtained, including both insertion loss and detection efficiency. The channel cross-talk level of the device was $-17.69$ dB. The dark count rate measured with a metal shielding cap on the fiber input terminal to prevent light insertion into the device was $< 10$ Hz throughout the investigated biasing range. Characterization of the temporal resolution in our detector system using a pulsed laser and an oscilloscope operated in histogram mode revealed a full width at half-maximum (FWHM) timing jitter of $\tau = (47.56 \pm 4.03)$ ps. The determined performance parameters enable multi-channel single-photon operation in the important telecommunication C-band, which may benefit future applications in wavelength-division multiplexed (WDM) quantum communication.

**SPS efficiency characterization**

A tunable, fiber-coupled NIR laser (New Focus TSL-6600 for the 1550 nm range / New Focus TLB-6700 Velocity for the 738 nm range) and a pair of calibrated attenuators (HP 8256A, 0-60 dB each) are used to generate a constant photon flux $\Phi$ entering the AWG. By measuring the optical power transmitted through the circuit's reference arm $P_{ref}$ using an optical power meter (HP 8163A) we precisely monitore the photon flux $\Phi$ inside the waveguide leading to the SPS. The count rate $R_c$ is measured with a specific flux $\Phi = 10^6 \ s^{-1}$ incident on the SPS and the dark count rate with disconnected fibers and metal shielding caps installed to prevent stray light insertion. The detection efficiency is obtained as the ratio of the dark count-corrected count rate and the incident photon flux, i.e. $\eta = (R_c - R_{dc})/\Phi$ .

**SPS timing jitter characterization**

The devices' timing jitter is determined using a low-jitter picosecond laser (PriTel FFL 40-M for NIR / ALS PiLas PiL044X for visible, timing jitter $< 2$ ps) and a fast digital sampling oscilloscope (Agilent Infiniium 54855A DSO) operated in histogram mode. Half of the laser light is heavily attenuated and routed toward the on-chip device, the other half toward a fast reference photodetector (New Focus 1611, timing jitter < 1 ps) which is installed outside the cryostat at room-temperature to generate a stable trigger signal. The electrical outputs obtained from the SNSPD and the reference detector are utilized as start and stop signals for the oscilloscope. The measured devices' timing jitter values are below 50 ps in both wavelength regimes. The timing jitter is device limited and independent of the wavelength range.

**Fluorescence imaging and lifetime mapping of silicon vacancy color centers in diamond**

Besides single-photon characterization in the NIR wavelength range, the broadband detection capability of the SNSPDs enables the implementation of advanced devices for visible light analysis. As an application example we investigate the low-intensity fluorescence signal of SiV color centers in diamond nanocrystals using our visible-range SPS. Diamond nanocrystals with incorporated SiV defect states are emerging as promising candidates for labelling in biological tissue. The hosting nanodiamonds can be easily functionalized and are biocompatible. In addition, the SiV emission in the deep red wavelength range complements genetic fluorophores, in particular the green-fluorescent protein (GFP) and the red-fluorescent protein (RFP). It exhibits high brightness and virtually no photobleaching (*36*, *37*). Moreover, our SPS circuit directly interfaces with the application of SiVs in single photon generation (*38*, *39*); it allows for suppressing the optical excitation wavelength on chip and provides further capabilities for spectral analysis in a scalable fashion.

The investigated nanodiamond cluster specimen is created by drop-casting a colloidal solution of nanocrystals onto a microscope slide. Clusters of nanocrystals form upon solvent evaporation. The scanning confocal microscope depicted in Fig. 1a is used for sample alignment, its excitation and the collection of the fluorescence light. The SiVs' zero phonon emission line (ZPL) is found to be centered at 738 nm and optical excitation is possible from 300 nm to 580 nm (see supplementary information).

We first analyze the emitted fluorescence spectrum using the eight-channel SPS upon continuous-wave (cw) excitation at 532 nm. The SPS's channel spacing is designed such that the 8-channel AWG covers the expected optical bandwidth of the SiV emission. The fluorescence signal is collected with a high-numerical aperture microscope objective (Zeiss EC Plan-Neofluar 100x/1.3NA) and coupled into an optical fiber after filtering out the excitation light using dichroic mirrors. The fiber is shielded from stray light with a metal coating and is used to route the light into our cryo-measurement platform for coupling into the photonic circuitry. The relative magnitudes of the measured count rates agree well with the reference spectrum obtained using a conventional spectrometer (Fig. 3), thus accurately reproducing the ZPL emission peak at 738 nm. By exploiting the apparatus' scanning capabilities, we are able to include spatial information and thus spectrally image the nanodiamond cluster. The cluster shape is well reproduced on each wavelength channel (see Fig. 4a).

Taking advantage of the SNSPDs high timing resolution and fast response time, in a complementary experiment, we replace the exciting 532 nm cw laser with a passively mode-locked laser which produces pulses of 32 ps duration (FWHM) at 440 nm wavelength (ALS PiLas PiL044X). In addition to the spectral image reported above, pulsed excitation allows for correlative imaging in a start-stop measurement: upon the emission of a pulse the laser triggers a start event in our TCSPC electronics (Picoquant Picoharp 300) and the stop signal is provided by the registration of the fluorescence photons by the SPS. The collection of multiple start-stop-time delay data points allows for the extraction of the specimen's fluorescence decay time (Fig. 4b). Such data is available on all channels, thus providing temporal as well as spectral information simultaneously. In combination with the scanning confocal microscope setup a lifetime map of the diamond cluster is obtained alongside the spectral information (Fig. 4c).

## Conclusions

Ultimately, the adoption of a quantum photonic approach to sensing and imaging holds the potential to herald a new level of experimental fidelity. By integrating multiple SNSPDs with wavelength-discriminating on-chip circuitry spectrally resolved single-photon detection with high timing precision can be realized in a single photonic device. The two circuit devices presented here serve as prototypes of a rich class of hybrid nanophotonic-superconducting systems. AWGs with tens of output waveguides provide room for large SNSPD arrays similar to CCD arrays in conventional spectrometers. Such devices could enable the live monitoring of numerous fluorophores over a wide spectral range. In particular, highly precise infrared and thermal imaging become possible owing to the SNSPDs' enormous spectral range.

The presented SPS operated in the telecommunication wavelength regime at 1550 nm hold promise for serving as receiver elements in wavelength-division multiplexed (WDM) optical networks with low signal strengths and, in particular, for quantum optical implementations which require the reliable detection of individual photons on tailored wavelengths. Owing to the SNSPDs' high temporal accuracy and good on-chip detection efficiency even more advanced protocols involving quantum cryptography and quantum key distribution on multiple wavelength channels and thus with high bandwidth are brought within reach.

## Acknowledgement

W. Pernice acknowledges support by DFG grant PE 1832/1-1 and by the Helmholtz Society through grant HIRG-0005. V. Kovalyuk, A. Korneev and G. Gol'tsman acknowledge financial support by RFBR grant 15-52-10044 and State Contract No. 14.B25.31.0007. We thank S. Diewald and S. Kühn for help with device fabrication as well as B. Voronov for help with NbN thin film deposition.

## Author contributions

W.P. conceived and supervised the research. O.K. fabricated the samples and carried out the measurements with the help of S.F., A.V. and V.K. V.K. deposited the NbN thin films with assistance from A.K. and G.G.. G.L. and C.N. prepared the nanodiamonds. All authors discussed the results and wrote the manuscript.

**References:**

1. J. Stuhler, Quantum optics route to market. *Nat. Phys.* **11**, 293–295 (2015).
2. A. Politi, M. J. Cryan, J. G. Rarity, S. Yu, J. L. O'Brien, Silica-on-Silicon Waveguide Quantum Circuits. *Science (80-. ).* **320**, 646–649 (2008).
3. M. Thompson, Quantum integration. *Nat. Photonics*. **8**, 160–160 (2014).
4. J. O'Brien, B. Patton, M. Sasaki, J. Vučković, Focus on integrated quantum optics. *New J. Phys.* **15**, 035016 (2013).
5. S. Tanzilli *et al.*, On the genesis and evolution of integrated quantum optics. *Laser Photonics Rev.* **6**, 115–143 (2012).
6. P. J. Shadbolt *et al.*, Generating, manipulating and measuring entanglement and mixture with a reconfigurable photonic circuit. *Nat. Photonics*. **6**, 45–49 (2011).
7. J. W. Silverstone *et al.*, On-chip quantum interference between silicon photon-pair sources. *Nat. Photonics*. **8**, 104–108 (2013).
8. F. Najafi *et al.*, On-chip detection of non-classical light by scalable integration of single-photon detectors. *Nat. Commun.* **6**, 5873 (2015).
9. C. Reimer *et al.*, Generation of multiphoton entangled quantum states by means of integrated frequency combs. *Science (80-. ).* **351**, 1176–1180 (2016).
10. A. J. Shields, Semiconductor quantum light sources. *Nat. Photonics*. **1**, 215–223 (2007).
11. N. Matsuda *et al.*, A monolithically integrated polarization entangled photon pair source on a silicon chip. *Sci. Rep.* **2**, 817 (2012).
12. M. J. Collins *et al.*, Integrated spatial multiplexing of heralded single photon sources. *Nat. Commun.* **4**, 2582 (2013).
13. W. H. P. Pernice *et al.*, High-speed and high-efficiency travelling wave single-photon detectors embedded in nanophotonic circuits. *Nat. Commun.* **3**, 1325 (2012).
14. A. Divochiy *et al.*, Superconducting nanowire photon-number-resolving detector at telecommunication wavelengths. *Nat. Photonics*. **2**, 302–306 (2008).
15. R. W. Heeres, L. P. Kouwenhoven, V. Zwiller, Quantum interference in plasmonic circuits. *Nat. Nanotechnol.* **8**, 719–722 (2013).
16. C. Schuck *et al.*, Quantum interference in heterogeneous superconducting-photonic circuits on a silicon chip. *Nat. Commun.* **7**, 10352 (2016).
17. O. Kahl *et al.*, Waveguide integrated superconducting single-photon detectors with high internal quantum efficiency at telecom wavelengths. *Sci. Rep.* **5**, 10941 (2015).
18. C. Schuck, W. H. P. Pernice, H. X. Tang, Waveguide integrated low noise NbTiN nanowire single-photon detectors with milli-Hz dark count rate. *Sci. Rep.* **3**, 1893 (2013).
19. T. Gerrits *et al.*, On-chip, photon-number-resolving, telecommunication-band detectors for scalable photonic information processing. *Phys. Rev. A*. **84**, 060301 (2011).

20. B. Calkins *et al.*, High quantum-efficiency photon-number-resolving detector for photonic on-chip information processing. *Opt. Express*. **21**, 22657 (2013).

21. E. Betzig *et al.*, Imaging Intracellular Fluorescent Proteins at Nanometer Resolution. *Science (80-. )*. **313**, 1642–1645 (2006).

22. A. Kirmani *et al.*, First-Photon Imaging. *Science (80-. )*. **343**, 58–61 (2014).

23. J. J. Schmied *et al.*, Fluorescence and super-resolution standards based on DNA origami. *Nat. Methods*. **9**, 1133–1134 (2012).

24. B. N. G. Giepmans, The Fluorescent Toolbox for Assessing Protein Location and Function. *Science (80-. )*. **312**, 217–224 (2006).

25. A. Kirchhofer *et al.*, Modulation of protein properties in living cells using nanobodies. *Nat. Struct. Mol. Biol.* **17**, 133–138 (2010).

26. K. Okabe *et al.*, Intracellular temperature mapping with a fluorescent polymeric thermometer and fluorescence lifetime imaging microscopy. *Nat. Commun.* **3**, 705 (2012).

27. J. W. Borst, A. J. W. G. Visser, Fluorescence lifetime imaging microscopy in life sciences. *Meas. Sci. Technol.* **21**, 102002 (2010).

28. P. Kapusta, M. Wahl, R. Erdmann, *Advanced Photon Counting* (Springer International Publishing, Cham, 2015; https://books.google.com/books?id=u6eqCAAAQBAJ&pgis=1), vol. 15 of *Springer Series on Fluorescence*.

29. D. J. S. Birch, D. McLoskey, A. Sanderson, K. Suhling, A. S. Holmes, Multiplexed time-correlated single-photon counting. *J. Fluoresc.* **4**, 91–102 (1994).

30. R. H. Hadfield, Single-photon detectors for optical quantum information applications. *Nat. Photonics*. **3**, 696–705 (2009).

31. M. D. Eisaman, J. Fan, a Migdall, S. V Polyakov, Invited review article: Single-photon sources and detectors. *Rev. Sci. Instrum.* **82**, 071101 (2011).

32. C. Schuck, W. H. P. Pernice, H. X. Tang, NbTiN superconducting nanowire detectors for visible and telecom wavelengths single photon counting on Si3N4 photonic circuits. *Appl. Phys. Lett.* **102**, 051101 (2013).

33. J. P. Sprengers *et al.*, Waveguide superconducting single-photon detectors for integrated quantum photonic circuits. *Appl. Phys. Lett.* **99**, 181110 (2011).

34. J. Lee *et al.*, Single-Molecule Four-Color FRET. *Angew. Chemie Int. Ed.* **49**, 9922–9925 (2010).

35. A. D. Hoppe, B. L. Scott, T. P. Welliver, S. W. Straight, J. A. Swanson, N-Way FRET Microscopy of Multiple Protein-Protein Interactions in Live Cells. *PLoS One*. **8** (2013), doi:10.1371/journal.pone.0064760.

36. Y.-R. Chang *et al.*, Mass production and dynamic imaging of fluorescent nanodiamonds. *Nat. Nanotechnol.* **3**, 284–288 (2008).

37. J. R. Maze *et al.*, Nanoscale magnetic sensing with an individual electronic spin in diamond. *Nature*. **455**, 644–647 (2008).

38. C. Wang, C. Kurtsiefer, H. Weinfurter, B. Burchard, Single photon emission from SiV

centres in diamond produced by ion implantation. *J. Phys. B At. Mol. Opt. Phys.* **39**, 37–41 (2006).

39. E. Neu *et al.*, Single photon emission from silicon-vacancy colour centres in chemical vapour deposition nano-diamonds on iridium. *New J. Phys.* **13**, 025012 (2011).

40. G. N. Gol'tsman *et al.*, Fabrication of nanostructured superconducting single-photon detectors. *IEEE Trans. Appiled Supercond.* **13**, 192–195 (2003).
16

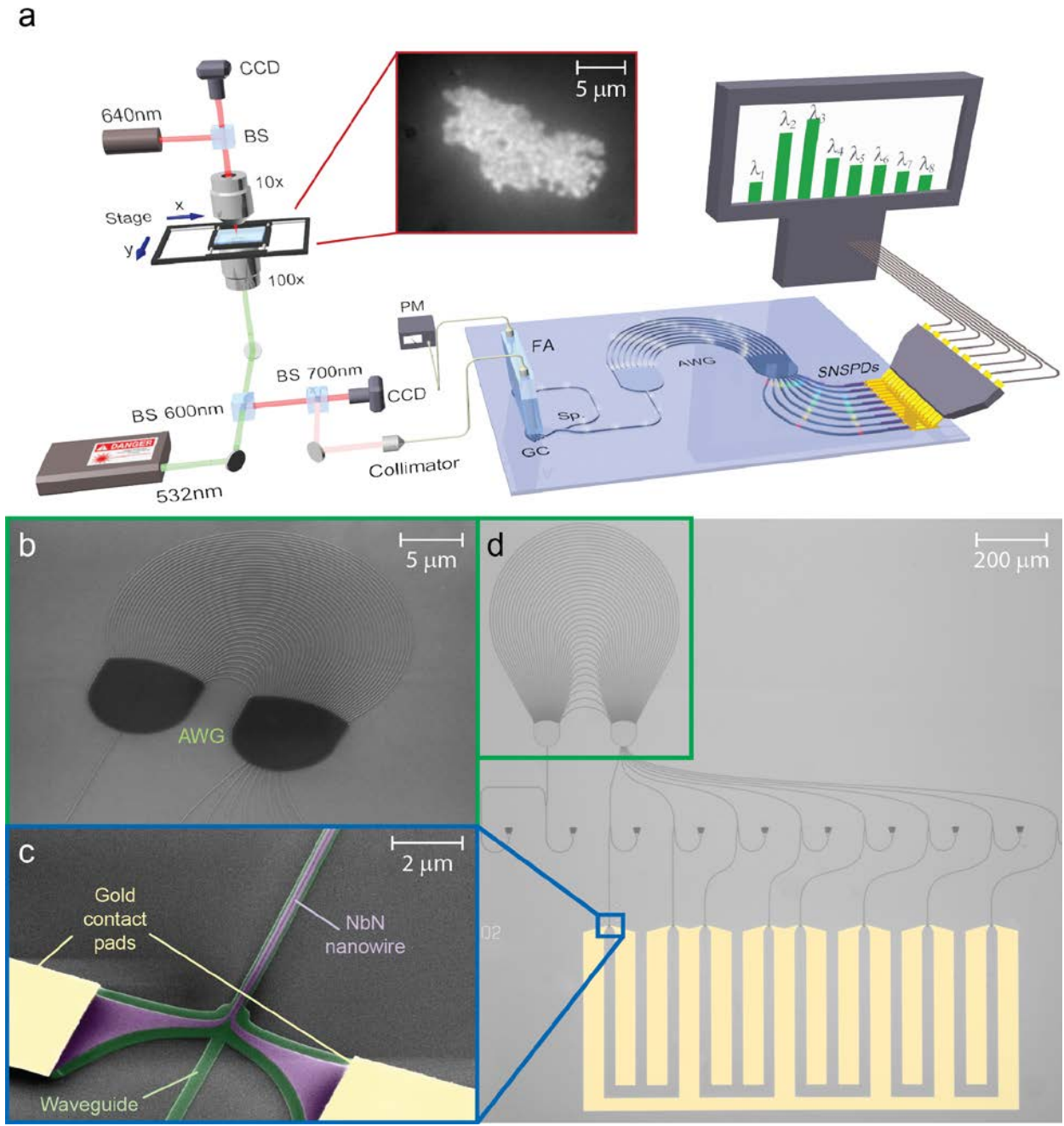

**Figure 1 | On-chip single-photon spectrometer.** (a) Illustration of the single photon spectroscopy setup including a confocal scanning system and the integrated single-photon spectrometer (SPS) chip which comprises an input grating coupler for light insertion and a reference port for output coupling, the AWG for spectral separation of the broadband input spectrum into eight output waveguides, and SNSPDs at the end of each output waveguide for single-photon detection. Inset: optical micrograph of a diamond nanocluster with embedded SiV color centers. Scalebar 5 μm. (b) Scanning electron micrograph of the AWG circuit section. (c) False color SEM micrograph of one of the 8 SSPDs. (d) Optical micrograph of one of the fabricated devices with nanophotonic elements and electrode array in false color.

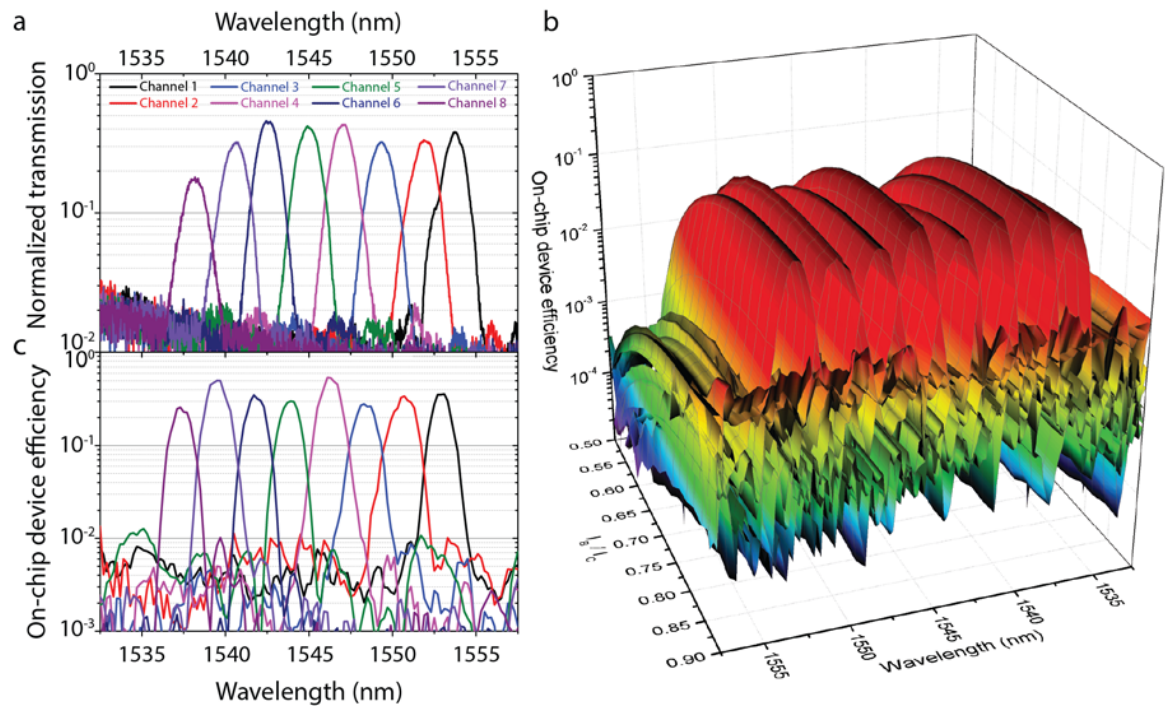

**Figure 2 | Performance of the single photon spectrometer at telecommunication wavelengths.** (a) Measured optical transmission in the telecommuniation C-band showing eight separate transmission channels. (b) Corresponding single-photon on-chip detection efficiency as a function of wavelength and bias current. (c) On-chip device efficiency as a function of wavelength biased at 90% of the SNSPDs' critical current.

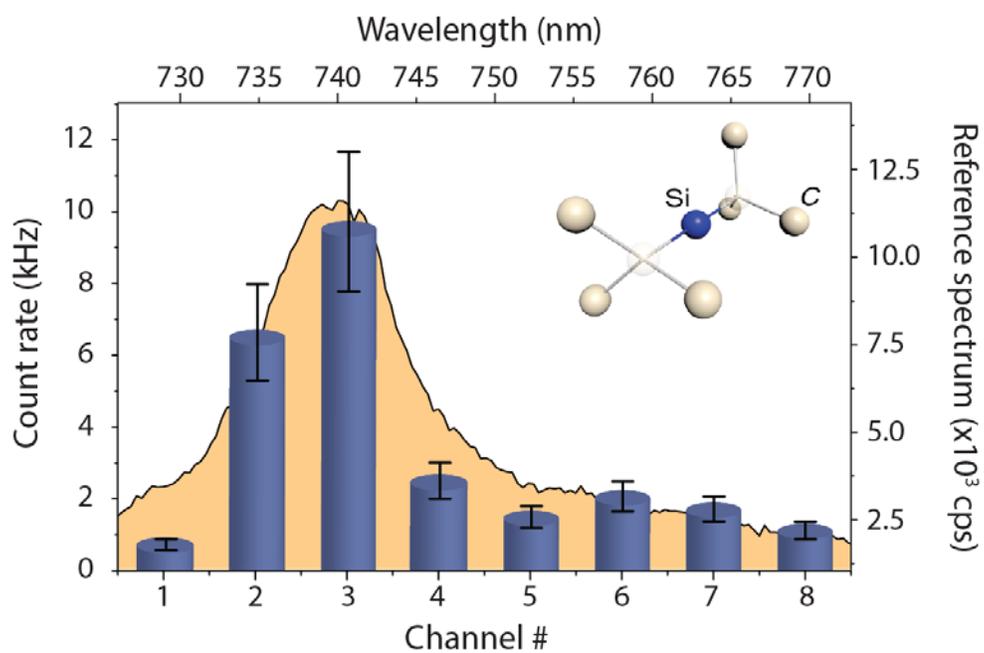

**Figure 3 | Single-photon spectroscopy of SiV color centers in diamond nanoclusters.** Emission spectrum of silicon vacancy color centers embedded in a diamond nanocluster recorded by an eight-channel SPS (blue bars) optimized for operation at 738 nm. Overlaid in yellow is the emission spectrum obtained with a conventional spectrometer. The inset shows a three-dimensional molecular model of the SiV color center.

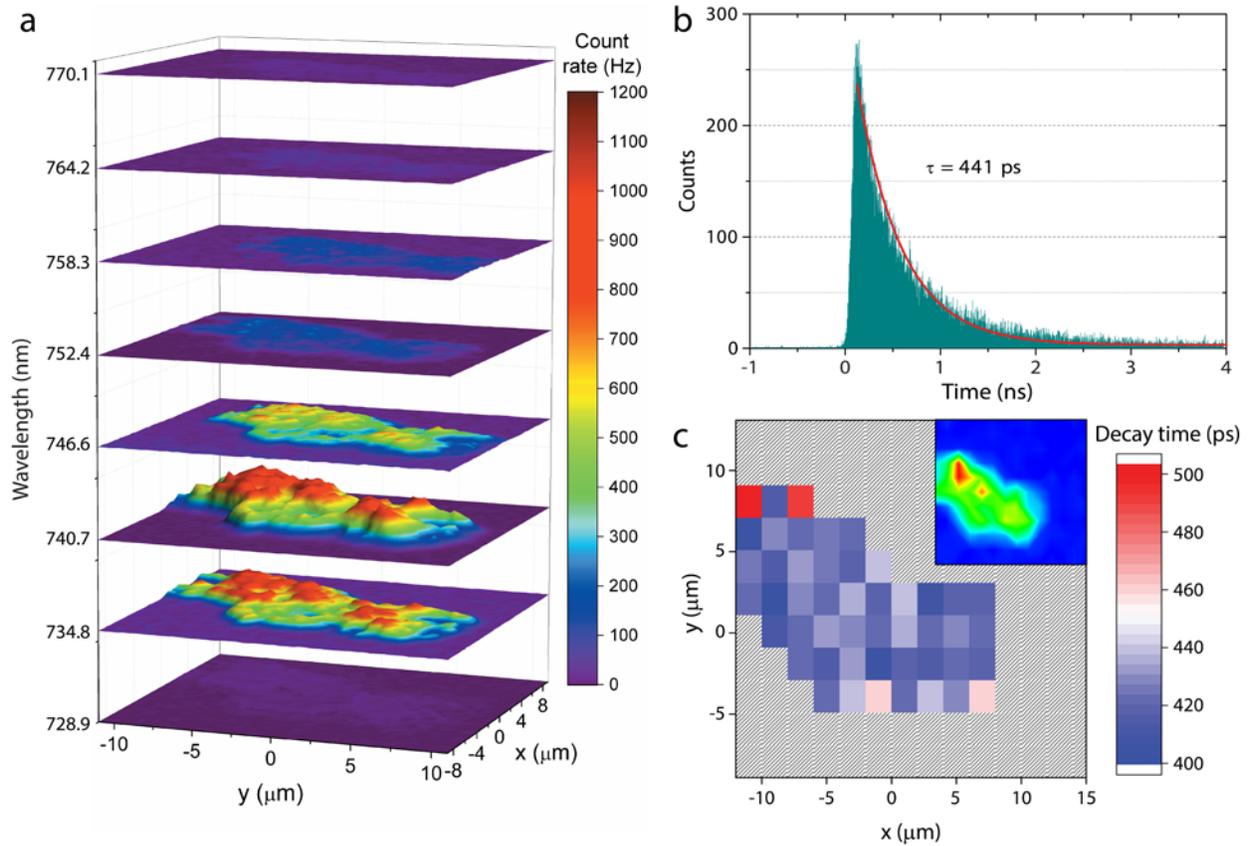

**Figure 4 | Confocal imaging with spectral and temporal single photon resolution.** (a) Spatial map of the count rate obtained in each of the 8 wavelength channels around 738 nm. (b) Single fluorescence decay trace of SiV centers obtained by TCSPC including single-exponential fit (solid red line). (c) Mapped out fluorescence decay times of the SiV centers found within the cluster area. Inset: confocal intensity scan of the same cluster.